\documentclass[runningheads]{llncs}
\usepackage{graphicx}
\usepackage{amsmath}
\usepackage{subfigure}
\usepackage{bbm}
\usepackage[ruled,vlined, linesnumbered]{algorithm2e}
\usepackage{array}
\usepackage{booktabs}
\usepackage{multirow}
\usepackage{graphicx} 
\usepackage{caption} 
\usepackage{float}
\usepackage{amsmath}
\begin{document}

\title{CoCoB: Adaptive Collaborative Combinatorial Bandits for Online Recommendation}
\titlerunning{Adaptive Collaborative Combinatorial Bandits}



\author {Cairong Yan\inst{1}\orcidID{0000-0003-0313-8833} \and 
Jinyi Han\inst{2} \thanks{corresponding author. E-mail:jinyihan099@gmail.com. \\This work is supported by the National Natural Science Foundation of China under Grant 62477006 and  62206046.}\orcidID{0009-0003-8380-2905} \and
Jin Ju\inst{1} \and
Yanting Zhang\inst{1}\orcidID{0000-0001-6317-1956} \and
Zijian Wang\inst{1}\orcidID{0000-0002-4096-9428} \and
Xuan Shao\inst{1}\orcidID{0000-0002-8147-5109}
}
\institute{School of Computer Science and Technology, Donghua University \and Shanghai Institute of Artificial Intelligence for Education, East China Normal University}

\maketitle              
\begin{abstract}

Clustering bandits have gained significant attention in recommender systems by leveraging collaborative information from neighboring users to better capture target user preferences. However, these methods often lack a clear definition of ``similar users'' and face challenges when users with unique preferences lack appropriate neighbors. In such cases, relying on divergent preferences of misidentified neighbors can degrade recommendation quality. To address these limitations, this paper proposes an adaptive \underline{Co}llaborative \underline{Co}mbinatorial \underline{B}andits algorithm (CoCoB). CoCoB employs an innovative two-sided bandit architecture, applying bandit principles to both the user and item sides. The user-bandit employs an enhanced Bayesian model to explore user similarity, identifying neighbors based on a similarity probability threshold. The item-bandit treats items as arms, generating diverse recommendations informed by the user-bandit's output. CoCoB dynamically adapts, leveraging neighbor preferences when available or focusing solely on the target user otherwise. Regret analysis under a linear contextual bandit setting and experiments on three real-world datasets demonstrate CoCoB’s effectiveness, achieving an average 2.4\% improvement in F1 score over state-of-the-art methods. The source code will be publicly available on GitHub.

\end{abstract}

%
%

\section{Introduction}\label{Section1}
\vspace{-0.2cm}
\begin{figure}[ht]
  \centering
\includegraphics[width=0.65\textwidth]{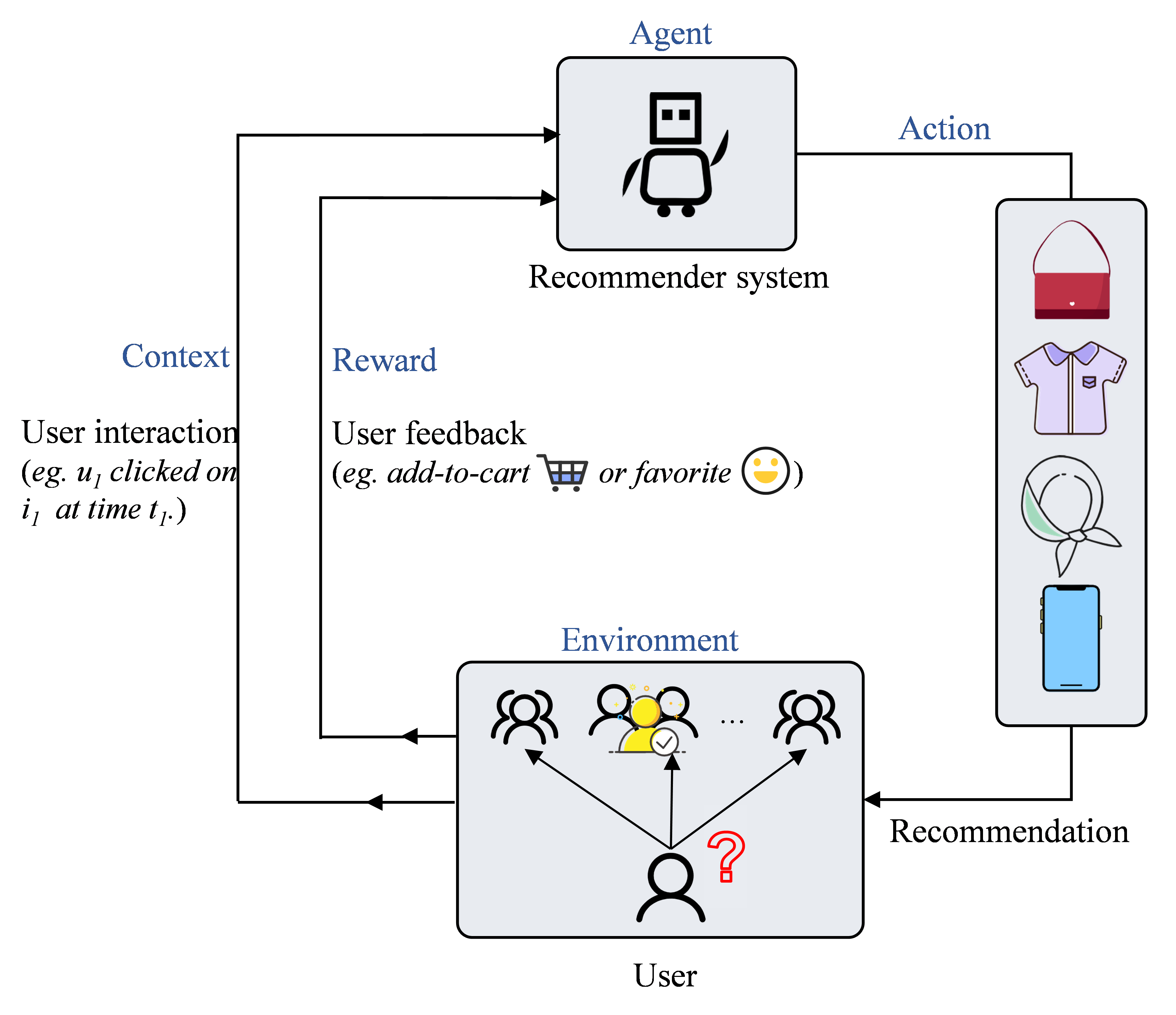}
\vspace{-0.2cm}

  \caption{A personalized recommendation task modeled as a clustering bandit problem. Black labels denote recommendation concepts, while blue labels represent bandit concepts. The \emph{Environment} box determines the target user's cluster.}
  \label{Figure1}
\end{figure}
\vspace{-0.2cm}

Personalized content recommendation is vital for online systems like e-commerce and streaming platforms \cite{he2020contextual}. Traditional methods, which assume fixed user preferences and item sets \cite{zhong2020best}, struggle in dynamic scenarios where preferences evolve and new users or items appear \cite{xu2020contextual}. This challenge stems from the exploration-exploitation (EE) dilemma: balancing the exploration of new items for long-term user satisfaction with exploiting known information for immediate recommendations. Multi-armed bandit (MAB), a reinforcement learning approach, effectively addresses the EE problem by sequentially selecting actions (arms) with unknown reward distributions, using user feedback to refine future choices.

Contextual MAB (CMAB) and combinatorial MAB, as extensions of MAB, are widely studied and applied in real-world recommender systems. CMAB incorporates contextual features for each arm, enabling more personalized and higher-quality decisions \cite{li2021constrained}. Combinatorial MAB selects multiple arms simultaneously, aligning better with practical needs by recommending multiple items at once. This paper leverages both frameworks to enhance the effectiveness of bandit algorithms in recommender systems.

Most bandit-based recommendation models, including CMAB and combinatorial bandits, take one of two extremes: a global bandit for all users \cite{2016Contextual}, which learns from shared reward information but may overlook individual preferences, or an independent bandit for each user \cite{lacerda2017multi}, offering fully personalized recommendations but struggling with sparse interaction data. To address this, some methods leverage user collaboration by clustering similar users based on shared preferences and using cluster information for recommendations \cite{2017Online,2018Learning}. This approach allows users within a cluster to collectively estimate arm rewards, as illustrated in Fig. \ref{Figure1}. However, for users with unique preferences, clustering fails to identify meaningful neighbors, making the process redundant or even detrimental to recommendation performance. This occurs because the preferences of ``neighboring users'' in such cases deviate significantly from the target user's preferences.


The clustering method is central to clustering bandits, with two main approaches. The first uses traditional methods like K-Nearest Neighbor (KNN) \cite{nguyen2014dynamic}, which are well-established but rely on static user attribute features, making it difficult to adapt to evolving preferences. The second employs graph-based methods, where nodes represent users, edges represent user similarities, and connected components define clusters \cite{gentile2014online,li2018online}. While this approach dynamically updates preferences by modifying edges, it fails to re-cluster similar users once separated, even if their preferences align later. Additionally, both approaches lack a clear quantitative definition of ``neighboring users''.

To deal with these issues in current clustering bandits approaches, this paper propose an adaptive \underline{C}ollaborative \underline{C}ombinatorial \underline{B}andits (CoCoB) algorithm. Built on a linear contextual bandit framework, CoCoB is a simple yet flexible solution. Our contributions are threefold:

\begin{itemize}
    \item Introduced a two-sided bandit framework, consisting of a user-bandit to model user similarities and an item-bandit to generate recommendation lists based on these preferences. This design enables the CoCoB algorithm to adaptively combine user collaboration and individual preferences.
    \item Proposed an enhanced Bayesian bandit-based strategy for selecting neighboring users. This approach quantitatively defines neighboring users by setting a similarity probability threshold, eliminating the need for user attributes. It effectively identifies and determines the presence of neighboring users. 
    \item Provided a thorough analysis of the CoCoB algorithm, including a theoretical regret bound under the linear contextual bandit framework, and experimental results on three real-world datasets showing CoCoB's superiority over state-of-the-art contextual bandit algorithms.
\end{itemize}


\begin{table}[ht]
  \caption{Notation information}
  \label{T_Notion}
  \centering
  \begin{tabular}{ll}
    \hline
    Notation & Explanation\\
    \hline
    $I$, $U$, $u$, $v$, $m$  & item set, user set, two users, $\left|U \right |=m$ \\
    $N_{ut}$             & neighboring user set of user $u$ at time $t$\\
    $\Bar N$        & average value of $|N_{ut}|$ \\
    $E_{t}$, $J$ 	    & candidate item set at time $t$, $\left|E_{t}\right|=J$\\
    $A_t$, $K$, $a$ & arm set at time $t$ $\left(A_t\subseteq E_t\right )$,  $\left|A_{t}\right|=K$, an arm\\
    $C_t$ & context vectors at time $t$ in the system\\
    $x_{a}$, $d$ & feature vector of arm $a$, vector dimension\\
    $\tilde{x}_{t}$ & expected context vector of $A_t$\\
    $\theta _u$, $\tilde{\theta}_{N_{ut}}$ & preference vector of $u$ or  $u$'s neighbors $N_{ut}$ \\
    $\mu_{at}$ & expected reward of arm $a$ at time $t$ \\
    $r_a$ & reward of arm $a$  \\
    $\Bar{r}_t$ & reward at time $t$ \\
    $r_t^\ast$ & reward of the optimal arms at time $t$ \\
    $\gamma$  & threshold of user similarity probability \\
    $\alpha$  & number of successes to find neighboring users\\
    $\beta$  & number of failures to find neighboring users \\
    $S$ &  interacted item set \\
    $Reg{(t)}$, $Reg(T)$  & cumulative regret at time $t$ or over $T$ time steps\\
    $p_{t}\left(v\mid u \right)$ & similarity probability distribution of users $u$ and $v$\\
    $M_{u}$, $\hat{M}_{N_{ut}}$ & estimated expected rewards for $u$ or $u$'s neighbors\\
    $w_{u}$, $\hat{w}_{N_{ut}}$ & feature matrix for $u$ or $u$'s neighbors \\
    $b_{u}$, $\hat{b}_{N_{ut}}$ & observed rewards for $u$ or $u$'s neighbors \\
    $B$ & optimal item set\\
    $e$     &exploration probability \\
    $\varepsilon$ & a constant, $0<\varepsilon<1$ \\
    $Z(a,T)$ & total numbers of arm $a$ is pulled over $T$ time steps\\
 \hline
\end{tabular}
\end{table}
%
%

\section{Related work}\label{Section2}
Significant progress has been made in the bandit problem, particularly with upper confidence bandit (UCB) \cite{auer2002finite} and Thompson sampling (TS) \cite{2011Analysis}. Variants such as clustering bandits\cite{gentile2014online,li2016collaborative,gentile2017context,2013The,wu2021clustering,jiang2020clustering} and combinatorial bandits \cite{ding2021hybrid,dong2022combinatorial} have been widely studied and applied in practical recommender systems.

\textbf{Clustering bandits}. While creating a unified bandit for all users or an individual bandit for each user has drawbacks, a more effective approach is to design bandits for groups of similar users. This approach leverages collaborative filtering, utilizing the preferences of neighboring users to improve recommendations. Gentile \emph{et al}. \cite{gentile2014online} introduced online clustering of bandits (CLUB), demonstrating the effectiveness of clustering users via a graph-based representation. SCLUB \cite{li2019improved} improved this by using sets to represent clusters, supporting split and merge operations. Li \emph{et al}. \cite{li2016collaborative} proposed collaborative filtering bandits (COFIBA), considering user and item interactions, where user clusters influence item clusters. Nguyen \emph{et al}. \cite{nguyen2014dynamic} applied k-means clustering on user preference vectors, presenting DynUCB, a dynamic clustering-based contextual bandit. Yan \emph{et al}. \cite{YAN2022109927} grouped users into clusters, treating those within the same cluster as neighbors to enhance recommendations. Gentile \emph{et al}. \cite{gentile2017context} further developed context-aware clustering of bandits (CAB), assuming users with similar expected rewards for the same item share preferences.

These methods focus on identifying neighboring users for the target user, but this process is time-consuming and lacks a quantitative or statistical definition of neighbors. While leveraging neighbors' preferences can improve recommendations, they fail when the target user has unique preferences and no true neighbors exist. In such cases, relying on misidentified ``neighbors'' can negatively impact recommendation accuracy. To balance user collaboration with individual preferences, this paper proposes the adaptive CoCoB algorithm, which effectively addresses both scenarios.

\textbf{Combinatorial bandits}. Many studies have extended MAB to combinatorial MAB, enabling multiple actions per round through various methods. For instance, the LMDH bandit \cite{ding2021hybrid} selects $K$ items by pulling arms $K$ times, offering diverse recommendations. The MP-TS algorithm \cite{komiyama2015optimal} extends TS to select the top $K$ arms from a candidate pool instead of a single arm. Additionally, approaches like \cite{chen2013combinatorial,qin2014contextual,dong2022combinatorial} employ super arms, a set of key arms following a specific distribution, and use semi-bandit feedback to observe their rewards.

The research focuses on recommending multiple items simultaneously to meet practical application needs. The top $K$ arms are selected from the candidates to form a recommendation list, with the average reward as the final outcome. In the CoCoB algorithm, the item-bandit is responsible for these functions. Table \ref{T_Notion} provides descriptions of the key terms used in this paper.

%
%

\section{Problem Formulation}\label{Section3}
We frame a typical e-commerce recommendation task as an adaptive collaborative combinatorial bandit problem, focusing on dynamically identifying neighboring users and integrating this into the combinatorial bandit framework.
\begin{figure*}[ht]
  \centering
  \includegraphics[width=0.7\linewidth]{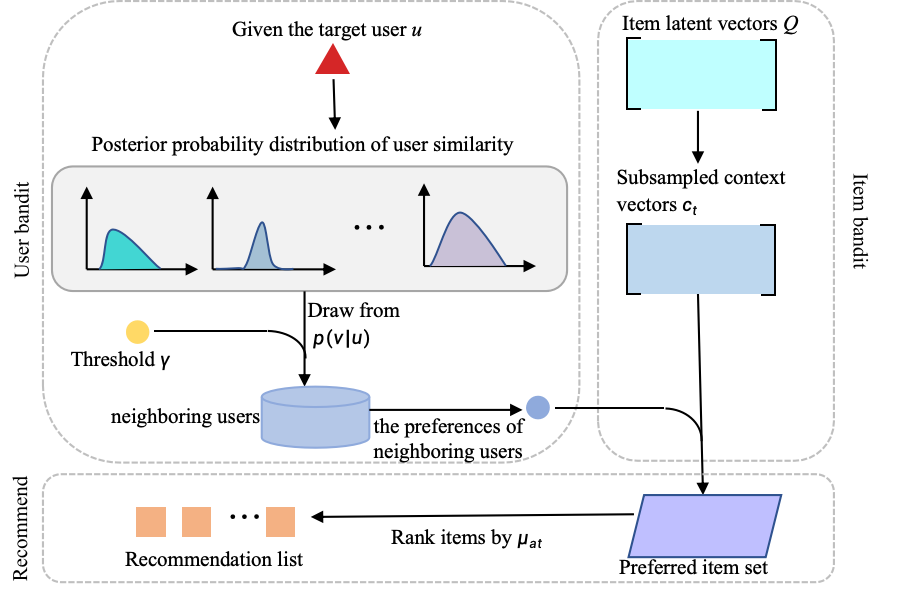}
  \caption{The general framework of the proposed CoCoB recommendation.}
  \label{Figure2}
\end{figure*}

\textbf{Environment}. Given an item set $I$ and a user set $U$ with $\left|U \right |=m$, at each iteration, a user $u$ is randomly selected. At time $t$, the agent identifies $u$'s neighbors, adding them to the neighboring user set $N_{ut}$. If neighbors are found, their preference data is incorporated to help infer $u$'s preferences. Since users' preferences evolve over time, the neighboring user set is updated regularly in each iteration.

\textbf{Actions}. At time $t$, the items in the candidate set $E_{t}\subseteq I$ are modeled as arms, where $\left|E_{t}\right|=J$. Based on the environment and reward experience from previous iterations, the agent makes selections from the top-$K$ performing arms tailored to the user, which can be expressed as $A_t \left(A_t\subseteq E_t \right )$.

\textbf{Context}. Contextual information, such as an item's color, can enhance action quality. At time $t$, the agent receives a set of context vectors $C_t= \left \{ x_1,\cdots ,x_J \right \}$ ($x_j\in R^d, 1\le j\le J$, where $d$ is the feature vector dimension), representing the contextual information of $J$ candidate items for the target user. Since the users change frequently and may be served multiple times, the contextual information the agent encounters is uncertain at each step.

\textbf{Policy}. In the contextual bandit setting, an unknown user vector $\theta _u \in R^d$ determines user $u$'s preferences. At time $t$, the expected payoff $\mu_{at}$ of item $a$ is typically expressed as a linear function of $\theta _u$.
\begin{equation}\label{equation 2}
    \mu_{at}= \tilde{\theta}_{N_{ut}}^{\mathrm{T}} x_a+f_{UCB}^a, 
\end{equation}
where $\tilde{\theta}_{N_{ut}}$ represents the context vector that determines the preferences of $N_{ut}$. $\mathrm{T}$ is the transpose symbol, and $f_{UCB}^a$ is a function that is used to represent the upper confidence bound of arm $a$. If there are no neighboring users, $N_{ut}=\left \{ u \right \} $. At time $t$, the $K$ arms with the highest estimated rewards are selected, and the responding items are added to the recommendation list $A_t$.

\textbf{Reward}. The reward reflects the target user's satisfaction with the recommended items, typically modeled as a Bernoulli distribution: a reward of 1 for satisfaction and 0 for dissatisfaction. In recommender systems, feedback can be explicit, such as ratings, or implicit, like clicks or add-to-carts. While explicit feedback is costly and requires user involvement, implicit feedback is more common and still indicates user preferences. At time $t$, the agent recommends $A_t$ and calculates the reward $\Bar{r}_t$ based on user feedback for the $K$ items in $A_t$. 
\begin{equation}\label{equation 2}
    \Bar{r}_t=\frac{1}{K} {\sum_{{a\in A_t}}}r_a, 
\end{equation}
where $r_a$ is the reward of arm $a$. The goal of the bandit algorithm is to maximize the total reward ${\textstyle \sum_{t=1}^{T}} \Bar{r}_t$ over $T$ steps. This paper primarily focuses on bounding the cumulative regret of the bandit algorithm. At time $t$,

\begin{equation}\label{equation 3}
    Reg\left(T \right)=\sum_{t=1}^{T} Reg(t) =\sum_{t=1}^{T} r_t^\ast -{r}_t,
\end{equation}
where $ Reg\left(T\right)$ is the cumulative regret over $T$ steps, and $r_t^\ast$ represents the reward of the optimal $K$ arms at time $t$.

The recommendation process proceeds as follows. At time $t$, the system observes the environment to check for neighboring users. If present, their preferences are incorporated into the context to aid in the recommendation. Based on this contextual information and prior knowledge, the system recommends a list of items to the target user. The system then collects feedback on the recommendations and calculates the reward to inform future recommendations.

%
%
\section{Methodology}\label{Section4}
We first present the proposed recommendation framework, followed by a description of the bandit algorithm, along with its complexity analysis and regret analysis.

\subsection{The proposed Recommendation Framework}
The general framework of the proposed CoCoB recommendation is illustrated in Fig. \ref{Figure2}. At time $t$, the system serves user $u$. Based on threshold $\gamma$, neighboring users are selected from the posterior probability distribution of user similarity for the target user (which may, in some cases, include only user $u$). Their preference information is then used to assist in the recommendation decision and generate the recommendation list.

\begin{algorithm}[H]
\SetAlgoLined
\KwIn{Parameters $\gamma$, $\alpha$, $\beta$.}
\KwOut{A recommendation list $A_t$ at time $t$.}
Initialize each $u$ by $b_{u} \gets 0 \in R^{d}$ and $M_{u} \gets I \in R^{d \times d}$\;
\For{$t \gets 1$ \KwTo $T$}{
    Initialize recommendation list $A_{t} \gets \emptyset$\;
    Receive a user $u$ to be served, $w_{u}\gets M_{u}^{-1}b_{u}$\;
    Find $u$'s neighborhood $N_{ut} \gets \emptyset$\;
    \For{all $v \in U$}{
        $p_{t} \left(v \mid u \right ) \gets Beta \left (\alpha \left (v \mid u \right ),\beta \left (v \mid u \right ) \right )$\;
        \If{$p_{t} \left (v \mid u \right ) \ge \gamma$}{
            $N_{ut} \gets N_{ut} \cup \left \{ v \right \}$\;
        }
    }
    \If{$N_{ut} = \emptyset$}{
        $N_{ut} \gets N_{ut} \cup \left \{ u \right \}$\;
    }
    Set $\hat{M}_{N_{ut}} \gets  \frac{1}{|N_{ut}|} {\textstyle \sum_{v\in N_{ut}}M_{v}}$,\\
    $\hat{b}_{N_{ut}} \gets \frac{1}{|N_{ut}|} {\textstyle \sum_{v\in N_{ut}}b_{v}}$,\\
    $\hat{w}_{N_{ut}}\gets \hat{M}_{N_{ut}}^{-1}\hat{b}_{N_{ut}}$\;
    Estimate $\mu_{at}$, $a \in E_{t}$\;
    Sort $\mu_{at}$ and select the top $K$ value with the order $(a_{1}^{t},\cdots ,a_{K}^{t})$\;
    Recommend $A_{t} \gets A_{t}\cup (a_{1}^{t},\cdots ,a_{K}^{t})$ to user $u$ and observe the reward $\left \{r_{a}\right \} _{a \in A_{t}}$\;
    Set $\tilde{x}_{t} = \frac{1}{K} {\textstyle \sum_{a \in A_{t}} x_{a}}$, \\
    $\Bar{r}_t = \frac{1}{K} {\textstyle \sum_{a \in A_{t}} r_{a}}$\;
    Run Update\;
}
\caption{CoCoB}
\label{Algorithm 1}
\end{algorithm}
\subsection{CoCoB Algorithm}
The CoCoB algorithm adaptively exploits the collaborative effect between users to extract user preferences. The pseudocode is shown in Algorithm \ref{Algorithm 1}. Three parts of CoCoB algorithm will be introduced below. 

\textbf{User-bandit part}. User-bandit implements the function of user collaboration by finding the neighboring users. As an improved TS bandit, it adaptively exploits user collaboration to discover user preferences. According to the TS principle, users are modeled as arms to explore similarities among users. There are available $m$ arms that can be pulled each time, representing that no more than $m$ users in the system can be neighboring users for the target user. At time $t$, for a target user $u$, user-bandit derives an estimate of $u$'s potential neighboring user $v$ from the Beta posterior with parameters $\alpha \left (v \mid u \right )$ and $\beta \left (v \mid u \right )$. $\alpha\left (v \mid u \right )$ represents the times $u$ is satisfied with the item selected by $v$ and $\beta \left (v \mid u \right )$ means the opposite. Here, $\alpha \left (v \mid u \right)= \alpha \left (u \mid v \right )$ and $\beta \left (v \mid u \right )= \beta \left (u \mid v \right )$. The sampled value $p_t \left (v \mid u \right )$ of each user $v$ given by the target user $u$ is defined as:
\begin{equation}
      p_t \left(v \mid u \right ) = Beta \left (  \alpha \left ( v \mid u \right ), \beta \left ( v \mid u \right ) \right ).
\end{equation}

Suppose $\gamma $ is a threshold parameter that controls the similarity probability of users. In this paper, if $p_t \left(v \mid u\right )\ge \gamma $, $v$ is defined as $u$'s neighboring user at time $t$. According to Bayesian theory, the larger $\gamma $ is, the more reliable the neighboring users are. $N_{ut}$ is the neighboring user set of $u$ at time $t$. If $N_{ut}$ is empty, it means that there are no neighboring users of $u$ under the constraint of the posterior probability threshold $\gamma $. To facilitate unified operation, user $u$ is added to $N_{ut}$ when $N_{ut}$ is empty. It represents that only $u$'s individual preference information is used to assist the recommendation. Moreover, the user preferences change dynamically. At different times, the same user may have different neighboring users or even no neighboring users. Therefore, the similarities between the target user and other users in the system need to be calculated each time. A detailed step-wise description of user-bandit is shown in lines 5$\sim $14 in Algorithm \ref{Algorithm 1}. 

\textbf{Item-bandit part}. Item-bandit is responsible for providing users with a list of recommendations for each round like other combinatorial bandits. First, each user profile is initialized at the beginning of Algorithm \ref{Algorithm 1}. Vector $w_{u}$ serves as a proxy to the unknown preferences of user $u$, which is defined as $w_{u}=M_{u}^{-1} b_{u}$. At time $t$, CoCoB servers user $u$ by providing an item list $A_t$ from a set of items $E_t$ represented as $C_t$. According to equation \ref{equation 2}, the upper confidence estimation $\mu_{at}$ of each arm $a$ at time $t$ is decided by aggregated confidence bounds and aggregated proxy vectors, defined as
\begin{equation}\label{equation 5}
         \mu _{at} = \hat{w}_{N_{ut}}^{\mathrm{T}} x_a+f \left(x_a \mid N_{ut} \right ) = \hat{w}_{N_{ut}}^{\mathrm{T}} x_a+e \sqrt{x_a^{\mathrm{T}} \hat{M}_{N_{ut}}^{-1} x_a\log \left ( 1+t \right ) }, 
\end{equation}
where $e$ is the exploration probability, and $\hat{w}_{N_{ut}}$ and $\hat{M}_{N_{ut}}$ are the parameters of $N_{ut}$, defined as line 15$ \sim $17 in Algorithm \ref{Algorithm 1}. CoCoB selects a set of items with the top $K$ largest upper confidence estimation according to equation \ref{equation 5}. A detailed stepwise description of the item-bandit approach is shown in lines 18$ \sim $20 in Algorithm \ref{Algorithm 1}.

\begin{algorithm}[H]
\SetAlgoLined
\LinesNumbered
\KwIn{Parameters $M_u, b_u, \tilde{x}_{t}, \Bar{r}_t$.}
\KwOut{$\alpha \left (v \mid u \right ), \beta \left (v \mid u \right )$ for all $v \in N_{ut}$.}
\textbf{Update $u$'s parameters:} \\
$M_{u} \gets  M_{u} + \tilde{x}_{t} \tilde{x}_{t}^{\mathrm{T}}$, $b_{u} \gets  b_{u} + \Bar{r}_t \tilde{x}_{t}$\;

\textbf{Update the similarity probability between $u$ and $v$ $(v \in N_{ut})$:} \\
\For{each $v \in N_{ut}$}{
    \eIf{$\Bar{r}_t > 0$}{
        $\alpha \left (v \mid u \right ) \gets \alpha \left (v \mid u \right ) + 1$\;
    }{
        $\beta \left (v \mid u \right ) \gets \beta \left (v \mid u \right ) + 1$\;
    }
}
\caption{Update Parameters}
\label{Algorithm 2}
\end{algorithm}
\textbf{Update part}. CoCoB observes the target user's response to the recommendation list $A_t$, calculates the reward obtained by each item, and finally takes the average reward $\Bar{r}_t$ of all items in $A_t$ as the recommendation reward and the average contextual vector $\tilde x_t$ of all items in $A_t$ as the feature presentation of the recommendation list. Then, CoCoB uses $\Bar{r}_t$ and $\tilde x_t$ to update the user parameter $w_u$ by solving a regularized least square problem as shown. Besides, it also updates the similarity probability distribution between target user and other users according to the recommendation reward $\Bar{r}_t$. If $\Bar{r}_t$ is greater than 0, it indicates that neighboring users in $N_{ut}$ provide positive help for this recommendation, then for each user $v$ in $N_{ut}$, update parameter $\alpha \left(v \mid u \right )$. Otherwise, update $\beta \left (v \mid u \right )$. The steps to update the parameters of user-bandit are shown in Algorithm \ref{Algorithm 2}.

\subsection{Theoretical Analysis}
\textbf{Algorithm complexity analysis}. Let $m$ be the number of users, $J$ the number of candidate items, and $d$ the feature dimension. At each step, finding neighboring users takes $O\left( m \right)$. Each recommendation has a complexity of $O\left(Jd^2\right)$, with matrix inverse updated using the Sherman–Morrison formula. Updating user parameters requires $O\left (d^2 \right )$, and updating the posterior probability distribution takes $O \left (\Bar N \right )$, where $\Bar N$ is the average number of neighboring users. Thus, the computational time complexity for $T$ rounds of CoCoB is $O\left (T \left (m+d^{2} \left (J+1 \right )+\Bar N \right )\right)$. In cases with no neighboring users, $\Bar N$ is 1.

\textbf{Regret analysis}. The design is primarily based on the proof idea of CAB \cite{gentile2017context}. The cumulative regret $Reg(T)$ of CoCoB consists of two terms, defined as:
\begin{equation}
    Reg(T)=Reg(T)_{\text {item-bandit }}+Reg(T)_{\text {user-bandit }}.
\end{equation}

The first term is the regret analysis of item-bandit, which follows the typical $\sqrt{T}$-style term seen in linear bandit regret analyses \cite{auer2002finite,li2010contextual}. For independent users, the first term takes the form $\sqrt{d \log T \sum_{t=1}^{T} m}$. However, in CoCoB, the dependence on the total number of users $m$ is replaced by a much smaller quantity, $\frac{m}{|N_{ut}|}$, which reflects the expected number of context-dependent user clusters. The regret bound for the item-bandit, ${Reg(T)}_{item-bandit}$, is defined as:
\begin{equation}\label{itemBanditRegret}
    \operatorname{Reg}(T)_{\text {item-bandit }} \leq 9 e \sqrt{d \log T \sum_{t=1}^{T} \frac{m}{|N_{ut}|}}.
\end{equation}

The second term is the regret analysis of user-bandit, a variant of the Multi-play MAB problem. In this case, the optimal arms are the top $B$ arms (i.e., arms $[B]$), while the suboptimal arms are the remaining ones (i.e., arms $N_{ut}\setminus[B]$). Let the selected suboptimal arm and the excluded optimal arm be $i$ and $j$, respectively. The regret is defined as
\begin{equation}
    \operatorname{Reg}(T)_{\text {user-bandit }}=\sum_{t=1}^{T}\left(\sum_{j \in[B]} \mu_{j}-\sum_{i \in N_{ut}} \mu_{i}\right).
\end{equation}

It was proven that for any strongly consistent algorithm and suboptimal arm $i$, the number of arm $i$ draws $Z(i, T+1)$ is lower-bounded as
\begin{equation}\label{armNum}
    {E}[Z(\{i, T+1\})] \le \frac{\log T}{(1-\varepsilon) \,\mathrm{dist}(\mu_i, \mu_B)},
\end{equation}
where $dist\left (p,q \right ) =p\log{\left ( {p}{q}\right )}+\left ( 1-p \right ) \log{{\left (1-p\right)}{\left (1-q\right)}}$ is the Kullback-Leibler ($KL$) divergence between two Bernoulli distributions with the expectation $p$ and $q$ and $\varepsilon$ $(0<\varepsilon<1)$ is a constant. For user-bandit, the highest similarity of target users is the users themselves, which means that $\max{\mu_{u\in U}}=1$. Then the loss in the expected regret at each round is given by
\begin{equation}
         \sum_{j \in[B]} \mu_{j}-\!\sum_{i \in N_{ut}} \mu_{i} =\!\!\!\!\sum_{j \in[B] \setminus \!\!\left[N_{ut}\right]}\!\!\!\mu_{j}-\!\!\!\sum_{i \in N_{ut} \setminus [B]}\!\! \!\mu_{i}\le \!\!\sum_{i \in N_{ut} \setminus[B]}\!\!\left(1-\mu_{i}\right).
\end{equation}

The regret is expressed as
\begin{equation}
    \begin{split}
        \operatorname{Reg}(T)_{\text {user -bandit }}\!&\le \sum_{t=1}^{T}\!\!\sum_{i \in N_{ut}\setminus [B]}\!\!\!\!\left(1-\mu_{i}\right)=\!\!\!\!\sum_{i \in N_{ut}\setminus[B]}\!\left(1-\mu_{i}\right) Z(i, T+1). \\
    \end{split}
\end{equation}

Combining the Equation \ref{armNum}, it is calculated that
\begin{equation}\label{userBanditRegret}
    \operatorname{Reg}(T)_{\text {user-bandit }} \leq \frac{\log T}{(1-\varepsilon )dist({{\mu }_{i}},{1})}\sum_{i \in N_{ut}\setminus[B]}\left(1-\mu_{i}\right).
\end{equation}

Therefore, according to Equation \ref{itemBanditRegret} and Equation \ref{userBanditRegret}, we can get the following Theorem 1. 

\textbf{Theorem 1.} \emph{
Suppose CoCoB is executed under the linear contextual bandit setting as described in Section 3, and the collaborative effect of users satisfy the assumptions stated in Section 4. Let $e= \mathcal{O}\left(\sqrt{logT} \right )$. Then the cumulative regret $ {\textstyle \sum_{t=1}^{T}} Reg\left (t \right )$ of CoCoB can be upper bounded as
}\\
\begin{equation}
    \begin{split}
    \operatorname{Reg} (T) \leq &9e \sqrt{d \log T \sum_{t=1}^{T} \frac{m}{|N_{ut}|}}+\frac{\log T}{(1-\varepsilon )dist({{\mu }_{i}},1)}\sum_{i \in N_{ut}\setminus[B]}\left(1-\mu_{i}\right).
     \end{split}
\end{equation}
%
%
\section{Experimental Result and Evaluation}\label{Experiments}
In this section, extensive experiments were conducted to answer the following questions:

RQ1: How does CoCoB perform on real-world datasets compared to the baseline algorithms for online recommendation tasks?

RQ2: How do critical parameters such as the recommended list length and probability threshold affect the performance of CoCoB?

RQ3: How does the efficiency of the CoCoB algorithm compare to that of the baseline algorithms?

\subsection{Experimental Settings}
\textbf{Datasets.} Experiments were conducted on three real-world datasets, IJCAI-15$\footnote[1]{\url{https://tianchi.aliyun.com/dataset/dataDetail?dataId=47}}$, Retailrocket$\footnote[2]{\url{https://www.kaggle.com/datasets/retailrocket/ecommerce-dataset}}$, and Yoochoose$\footnote[3]{\url{http://2015.recsyschallenge.com/challenge.html}}$. To standardize the experimental data, offline user data was selected with sequence lengths between 10 and 20, and 1,000 users were randomly sampled as test users. After preprocessing, IJCAI-15 contains 15,681 interaction records from 1,000 users and 8,348 items; Retailrocket includes 13,748 records from 1,000 users and 4,822 items; and Yoochoose has 13,602 records from 1,000 users and 2,619 items. Following previous work \cite{2021Modeling}, user preferences for interacting items were assigned ratings: 1 for clicks, 2 for favorites, 3 for add-to-cart, and 4 for purchases. Matrix factorization was used to generate the users' contextual feature vectors.

\textbf{Evaluation metrics.} Four metrics were used to measure the quality and effectiveness of the algorithms: cumulative reward (CR), precision, recall, and F1 score. CR represents the total rewards achieved by the algorithm, with higher values indicating closer alignment to the optimal strategy and better performance. The F1 score evaluates the balance between precision and recall across all evaluation rounds. Detailed definitions of these metrics are provided below.

\begin{align}
     & precision=\frac{1}{T} {\textstyle \sum_{t=1}^{T}} \frac{\left|A_t\cap S \right |}{K}, \quad recall=\frac{1}{T} {\textstyle \sum_{t=1}^{T}}
    \frac{\left|A_t\cap S \right|}{\left|S\right|},
\end{align}

\begin{align}
    & F1=\frac{1}{T} {\textstyle \sum_{t=1}^{T}} \frac{2\times precision\left(t\right)\times recall\left(t\right)}{precision\left(t\right) + recall\left(t\right)}, 
\end{align}
where $A_t$ is the recommendation item set at round $t$, $S$ represents the item set interacted with the user in the test data, and $precsion\left(t\right)$ and $recall\left(t\right)$ denote the recommendation precision and recall at time $t$.

The \emph{replayer} method was applied to assess the performance of the algorithms across the three aforementioned datasets. It operates by leveraging historical logs, where these logs are presumed to be the outcome of random recommendation processes. The core concept behind the \emph{replayer} method is to replay each user interaction recorded in the historical logs within the context of the evaluated algorithm. A critical criterion for a successful match between historical recommendations and the testing algorithm is when the item recommended by the bandit algorithm aligns with the one contained in the historical log. 

\textbf{Baselines.} Deep learning recommendation algorithms are primarily employed in offline recommendation tasks, capitalizing on abundant training data. In contrast, Bandit algorithms excel in online learning environments, specifically within the realm of exploration-exploitation scenarios. Given this distinction in application domains, the comparative analysis will focus on assessing CoCoB against the following baseline bandit algorithms: $C^2$UCB \cite{qin2014contextual}, $DC^3MAB$ \cite{YAN2022109927}, CAB \cite{gentile2017context}, $K$-LinUCB \cite{li2010contextual}, VarUCB \cite{xu2016dynamic}, and TV-UCB \cite{yan2023thompson}.

\textbf{Implementation.} The experiments were conducted on an Intel(R) Xeon E5-2620 v4 server (8-core 2.10GHz) with 256GB RAM, using python 3.7. All algorithms share common parameters: 50 arms, a recommendation list length of $K=10$, and random exploration probability $e=0.1$. Feature vector dimension were set to 6, 16, and 25 for IJCAI-15, Retailrocket, and Yoochoose, respectively. In addition, DynUCB used 10 user clusters, while CoCoB was configured with $\gamma=0.8$, $\alpha=15$, and $\beta=15$.

\subsection{Comparison Against Baselines (RQ1)} 





\begin{figure*}[ht]
  \centering
\includegraphics[width=1\linewidth]{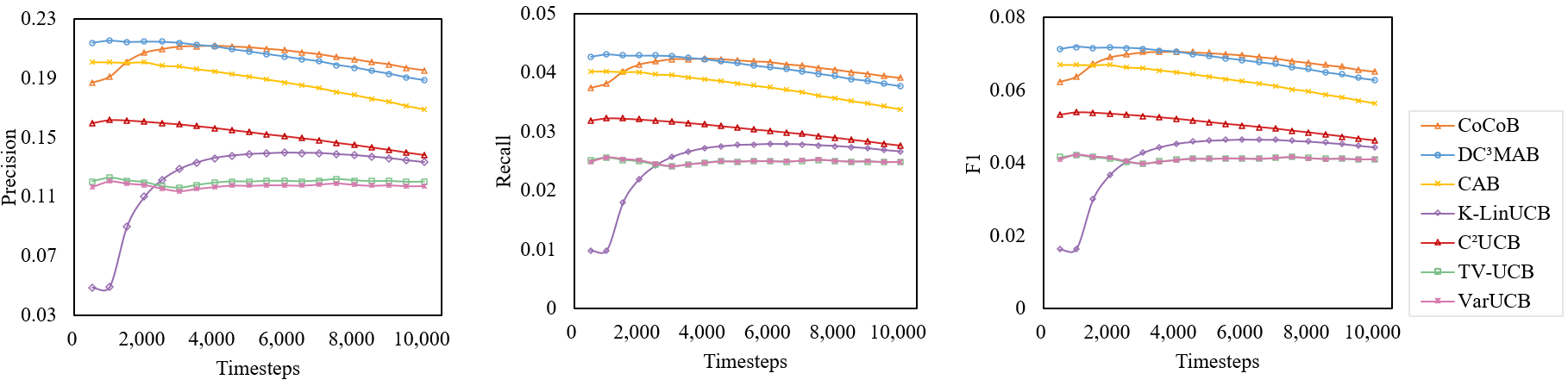}
  \caption{Recommendation performance of different algorithms on IJCAI-15.}
  \label{IJCAI}
\end{figure*}
\begin{figure*}
  \centering
\includegraphics[width=1\linewidth]{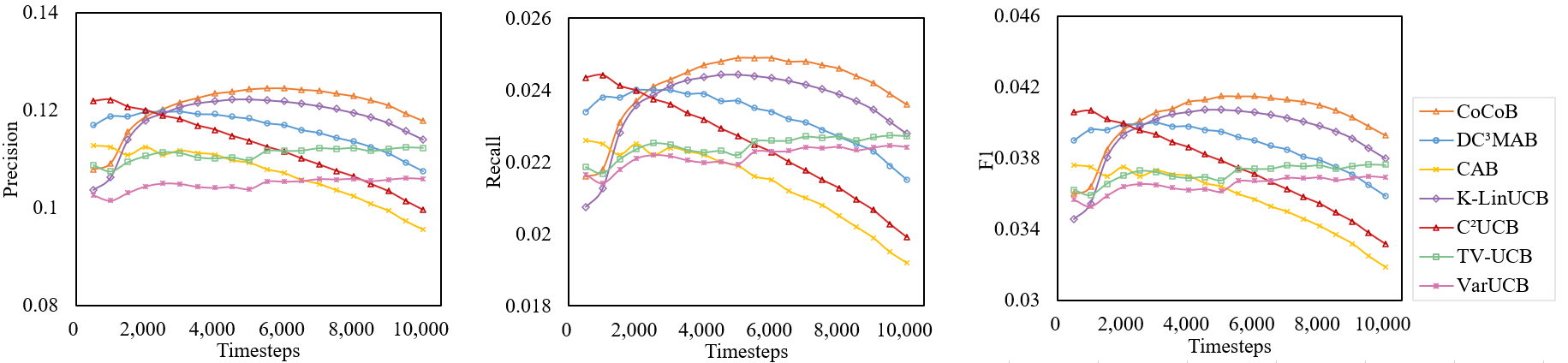}
  \caption{Recommendation performance of different algorithms on Retailrocket.}
  \label{Retailrocket}
\end{figure*}
\begin{figure*}
  \centering  \includegraphics[width=1\linewidth]{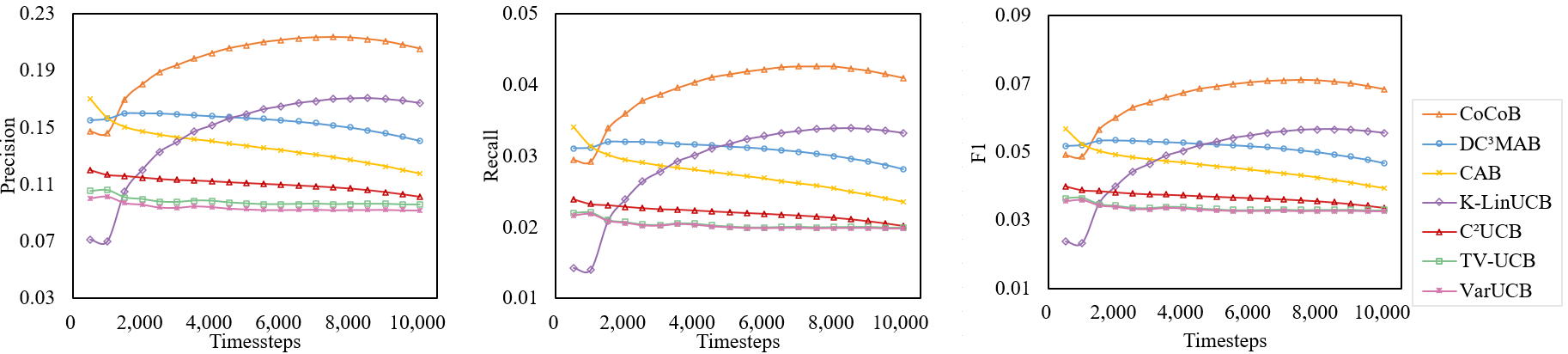}
  \caption{Recommendation performance of different algorithms on Yoochoose.}
  \label{Yoochoose}
\end{figure*}

Fig. \ref{IJCAI}, Fig. \ref{Retailrocket}, and Fig. \ref{Yoochoose} show the comparison of different performance metrics of different algorithms on the three datasets. All results were averaged over five runs. From the results, we can get the following observations.

First, when the number of experiment rounds exceeds 2,000, CoCoB outperforms other algorithms on all metrics. However, when the Timesteps value is small, CoCoB performs poorly compared to other baselines, such as $DC^3MAB$. This is because, for new users, the system lacks information about their preferences and assumes that similar users do not exist in the system. In such cases, CoCoB relies solely on the target user's information to make recommendations. In contrast, other baselines consistently utilize information from other users to assist with recommendations. Moreover, a common pattern emerges in the performance of these algorithms, as shown in the three figures: an initial dip in recommendation accuracy is followed by a subsequent increase that eventually stabilizes. This phenomenon can be attributed to the exploratory phase of bandit algorithms, during which they gather user preference data, temporarily reducing accuracy. Over time, these algorithms increasingly rely on the established preferences, leading to improved and stable performance.

Second, on the Yoochoose dataset, the recommendation performance of these baselines varies significantly with the increasing number of recommendation rounds. On the other two datasets, the performance is relatively stable. The main reason is that the behaviors in the sequences selected on the Yoochoose dataset have a considerable period, the bandit algorithms are more sensitive to time, and the noise information brought by these long-term historical data degrades the recommendation performance.

\begin{figure*}[htb]
  \centering
  \includegraphics[width=1\linewidth]{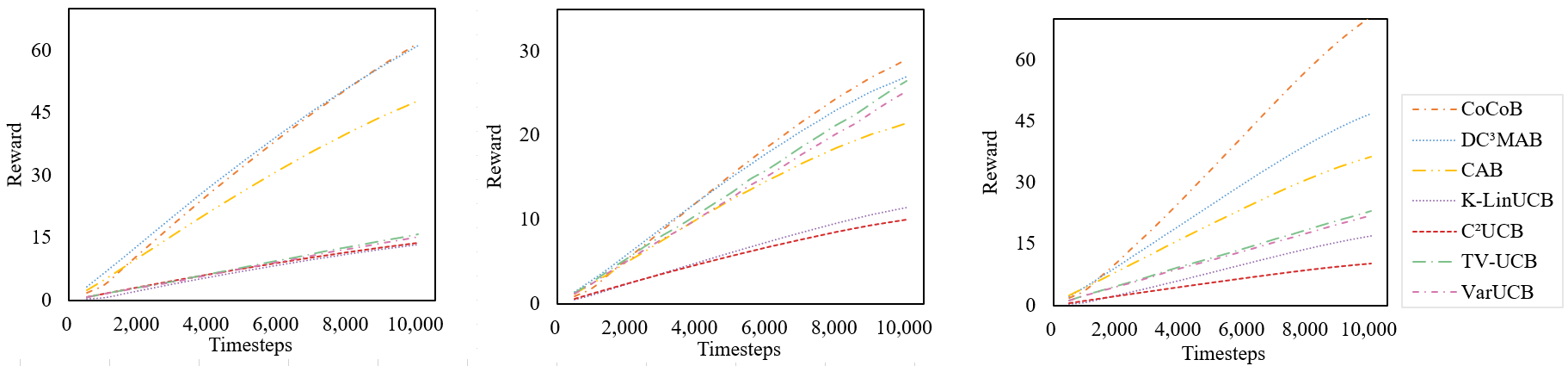}
  \caption{The performance of different algorithms on the IJCAI-15, Retailrocket, and Yoochoose datasets as measured by cumulative reward. The reward values correspond to the experimental results divided by 1,000.}
  \label{Reward}
\end{figure*}

\subsection{Analysis of Important Parameters (RQ2)}
This experiment examines the influence of critical parameters in the CoCoB algorithm, specifically assessing the impact of $K$ on recommendation list length within the combined bandits approach and $\gamma$ on the relevance of similar users. All other parameters are configured as in RQ1. The results, averaged over five runs of 10,000 epochs each, are presented in Table \ref{Table 1}, yielding the following observations.

\textbf{Effect of $K$}. Increasing $K$ consistently enhances performance metrics, with a more significant improvement observed when raising $K$ from 1 to 5 compared to 5 to 10. This indicates that the number of hits does not scale linearly with the recommendation list length. Overall, a suitable increase in $K$ improves both diversity and accuracy.

\textbf{Effect of $\gamma$}. Experimental results reveal a positive correlation between $\gamma$ and recommendation performance, with optimal performance observed at $\gamma =0.8$ across all three datasets. These findings highlight the role of user similarity in the datasets and offer guidance for parameter tuning. When user similarity is low, setting $\gamma$ closer to 1, while a value nearer to 0 is preferable for higher user similarity. Overall, $\gamma$ provides flexibility in fine-tuning the similarity degree among neighboring users, enabling the algorithm to adapt to diverse scenarios.

\begin{table*}
      \caption{ Effects of $K$ and $\gamma$ on recommendation performance (F1 score)}
  \label{Table 1}
  \centering
  \resizebox{\textwidth}{!}{
  \begin{tabular}{ccccccccccccccc}
    \hline
      \multirow{2}{*}{\textbf{Dataset}} &\multicolumn{4}{c}{\textbf{IJCAI-15}} & &\multicolumn{4}{c}{\textbf{Retailrocket}} & &\multicolumn{4}{c}{\textbf{Yoochoose}} \\
      \cline{2-5} \cline{7-10} \cline{12-15}
      &\textbf{min} &\textbf{max} &\textbf{mean} &\textbf{std} &   &\textbf{min} &\textbf{max} &\textbf{mean} &\textbf{std} &   &\textbf{min} &\textbf{max} &\textbf{mean} &\textbf{std}\\
    \hline
    $K=1\left ( \gamma =0.8 \right )$	
    &0.0145	&0.0148	&0.0147	&0.0001	 &	&0.0099	&0.0109	&0.0106	&0.0003	&	&0.0172	&0.0178	&0.0175	&0.0002\\
   $K=5 \left ( \gamma =0.8 \right )$	&0.0434	&0.0439	&0.0436	&0.0002	&	&0.0275	&0.0282	&0.0277	&0.0003	&	&0.0507	&0.0519	&0.0512	&0.0004\\
   $K=10 \left ( \gamma =0.8 \right )$	&0.0647	&0.0655	&0.0651	&0.0002	&	&0.0390	&0.0394	&0.0392	&0.0001	&	&0.0673	&0.0685	&0.0680	&0.0004\\
    $K=15 \left ( \gamma =0.8 \right )$	&0.0804 	&0.0809 &0.0806 &0.0002 &	&0.0485 &0.0490 	&0.0487 &0.0002 &	&0.0782 &0.0791 &0.0787 	&0.0003 \\
    $K=20 \left ( \gamma =0.8 \right )$	&0.0923 	&0.0929 &0.0926 &0.0002 &	&0.0561 	&0.0565 	&0.0564 	&0.0001 	&	&0.0867 	&0.0874 	&0.0871 	&0.0002 
\\
    $K=25 \left ( \gamma =0.8 \right )$	&0.1022 	&0.1028 	&0.1025 	&0.0002 	&	&0.0618 	&0.0622 	&0.0620 	&0.0002 	&	&0.0929 	&0.0935 	&0.0931 	&0.0002 
\\
    $K=30 \left ( \gamma =0.8 \right )$	&0.1106 	&0.1114 	&0.1109 	&0.0003 	&	&0.0662 	&0.0665 	&0.0664 	&0.0001 	&	&0.0974 	&0.0975 	&0.0975 	&0
\\
\hline
$\gamma =0.1\left ( K=10 \right )$	&0.0596	&0.0614	&0.0604	&0.0007	&	&0.0332	&0.0351	&0.0341	&0.0007		& &0.0433	&0.0452	&0.0443	&0.0007\\
$\gamma =0.2\left ( K=10 \right )$	&0.0607	&0.0619	&0.0611	&0.0005	&	&0.0335	&0.0346	&0.0341	&0.0004		& &0.0438	&0.0463	&0.0447	&0.0008\\
$\gamma =0.3 \left ( K=10 \right )$	&0.0604	&0.0629	&0.0613	&0.0008	&	&0.0338	&0.0346	&0.0343	&0.0003		& &0.0431	&0.0472	&0.0449	&0.0015\\
$\gamma =0.4\left ( K=10 \right )$	&0.0608	&0.0617	&0.0613	&0.0004	&	&0.0345	&0.0356	&0.0348	&0.0004		& &0.0441	&0.0462	&0.0451	&0.0007\\
$\gamma =0.5\left ( K=10 \right )$	&0.0602	&0.0624	&0.0617	&0.0008	&	&0.0341	&0.0361	&0.0354	&0.0007		& &0.0456	&0.0467	&0.0461	&0.0004\\
$\gamma =0.6\left ( K=10 \right )$	&0.0615	&0.0628	&0.0624	&0.0004	&	&0.0353	&0.0359	&0.0357	&0.0002		& &0.0471	&0.0485	&0.0477	&0.0005\\
$\gamma =0.7\left ( K=10 \right )$	&0.0622	&0.0626	&0.0625	&0.0001	&	&0.0365	&0.0370	&0.0368	&0.0002		& &0.0524	&0.0529	&0.0526	&0.0002\\
$\gamma =0.8\left ( K=10 \right )$	&0.0648	&0.0656	&\textbf{0.0651}&0.0003	&	&0.0390	&0.0396	&\textbf{0.0393}	&0.0002		& &0.0678	&0.0688	&\textbf{0.0683}	&0.0003\\
$\gamma =0.9\left ( K=10 \right )$	&0.0646	&0.0651	&0.0648	&0.0002	&	&0.0386	&0.0390	&0.0387	&0.0002	&	&0.0668	&0.0673	&0.0670	&0.0002\\
\hline 
\end{tabular}
}
\end{table*}

\subsection{Efficiency Comparison (RQ3)}
Table \ref{Table 2} summarizes the total running times (in minutes) for CoCoB and the baseline algorithms, averaged over five random runs. Among the evaluated algorithms, $C^2$UCB achieves the fastest performance by utilizing a single parameter across all users, resulting in time savings during parameter updates. In comparison, CoCoB exhibits notable computational efficiency compared to similar clustering bandit algorithms, such as DynUCB and CAB. This efficiency stems from CoCoB's adaptive strategy for identifying neighboring users, whereas DynUCB relies on the conventional KNN method for assessing user similarity. CAB, in contrast, exhibits the slowest execution time due to its intricate approach to selecting neighboring users. Its time complexity is determined to be $O\left (mJ \right )$, making it computationally intensive. Furthermore, in the context of round $T$, the time complexity expands to $O\left(T\left(mJ+d^{2} \left(J+1\right)+\Bar N\right)\right)$, significantly surpassing that of CoCoB.

\begin{table}
  \caption{Comparison of the total running time of different algorithms (m)}
  \label{Table 2}
  \centering
  \begin{tabular}{ccccccc}
    \hline
     \multicolumn{2}{c}{\textbf{Dataset}} & &\multicolumn{4}{c}{\textbf{Running time}}\\
     \cline{1-2} \cline{4-7}
      Name &Length & &CAB &DynUCB &$C^2$UCB &CoCoB  \\
    \hline
   IJCAI-15	 &15,681	& &1,292	&989	&87	&128\\
Retailrocket 	&13,748	& &1,325	&894	&67 &140	\\
Yoochoose 	&13,602 &	&1,418	&836 &62 &115\\	
 \hline
\end{tabular}
\end{table}
%
%
\section{Conclusion}\label{Conclusion}
This paper proposes CoCoB, a novel bandit algorithm that enhances online recommendation performance by leveraging user collaboration. CoCoB adopts a dual mechanism with user-bandit and item-bandit components. The user-bandit, an improved Thompson sampling model, dynamically identifies neighboring users based on a predefined similarity threshold, ensuring alignment with the target user's preferences. The item-bandit, a combinatorial bandit, generates a recommendation list in each iteration. Empirical results demonstrate its superior performance and efficiency. Future work will explore the sharp regret boundary of CoCoB to further validate its theoretical underpinnings.



\bibliographystyle{unsrt}
\bibliography{CoCoB}

\end{document}